\pdfoutput=1

\documentclass[11pt]{article}

\usepackage[preprint]{acl}

\usepackage{times}
\usepackage{latexsym}

\usepackage[T1]{fontenc}

\usepackage[utf8]{inputenc}

\usepackage{microtype}

\usepackage{inconsolata}

\usepackage{graphicx}

\usepackage{amsmath}

\title{The Echoes of the 'I': Tracing Identity with Demographically Enhanced Word Embeddings}

\author{Ivan Smirnov \\
  University of Technology Sydney \\
  \texttt{ivan.smirnov@uts.edu.au}}

\begin{document}
\maketitle
\begin{abstract}
Identity is one of the most commonly studied constructs in social science. However, despite extensive theoretical work on identity, there remains a need for additional empirical data to validate and refine existing theories. This paper introduces a novel approach to studying identity by enhancing word embeddings with socio-demographic information. As a proof of concept, we demonstrate that our approach successfully reproduces and extends established findings regarding gendered self-views. Our methodology can be applied in a wide variety of settings, allowing researchers to tap into a vast pool of naturally occurring data, such as social media posts. Unlike similar methods already introduced in computer science, our approach allows for the study of differences between social groups. This could be particularly appealing to social scientists and may encourage the faster adoption of computational methods in the field.
\end{abstract}

\section{Introduction}
Identity is central and one of the most commonly studied constructs in the social sciences, shaping our understanding human behaviour, and society more generally \cite{leary2003self}. While there is no universally accepted definition of identity, it generally refers to individual's self-perception that consists of self-ascribed personal traits, beliefs about themselves, as well as self-categorization into particular social groups and roles.

Research on identity spans disciplines from psychology to sociology, and from linguistics to political science  offering rich theoretical insights into identity \cite{vignoles2011introduction}. However, measuring identity and related constructs remains challenging, which is why there is still a clear need for empirical studies that would allow to validate and refine existing theories \cite{mclean2015field}. Established methods typically require the annotation of survey data by experts who have to be specially trained. Take, for instance, Loevinger theory of ego development \cite{loevinger1976ego} which is generally considered as one of the most empirically supported theories of personality development \cite{gilmore2001critical}. Traditionally, ego development is measured via the Washington University Sentence Completion Test (WUSCT) \cite{hy1996measuring}. That is a projective technique where participants are asked to complete sentence stems such as "What gets me in trouble..." or "A girl has a right to...". While WUSCT has been shown to be a reliable and valid method of measuring ego development \cite{gilmore2001critical}, its administration is resource-intensive and requires a specialized training for raters. At the same time, recent developments in computational methods suggest that social media data, at least at a macro level, can aid in assessing psychological constructs \cite{pellert2022validating}. This could pave the way for alternatives to traditional survey-based assessments. 

Computational approaches and natural language processing have been previously applied to study identity. In particular LIWC \cite{tausczik2010psychological} -- a popular dictionary-based method -- has been used to analyze responses to WUSCT \cite{lanning2018personality} or to identify salient identity in social media posts \cite{koschate2021asia}. In our work, we propose using word embeddings as they allow capturing more complex semantic relationships in the text by considering the context in which words are used.

The common approach to using word embeddings in social science is to consider projections on semantic axes in word-vector space. It has been previously demonstrated that this technique could effectively recover human sentiments, judgments, and perceptions \cite{an2018semaxis,grand2022semantic}. This enabled computational social scientists to extract insights from large text corpora. The potential of this approach was most notably demonstrated in studies on stereotypes \cite{garg2018word,boutyline2023school}. 

Typically, a word embedding model is trained on a specific corpus of interest. Then, the distance between target words and predefined reference poles, represented by opposing words or sets of words, is considered. This distance is interpreted as the semantic closeness between target words and reference poles, providing insights into underlying associations and relationships. More concretely, it has been shown that certain occupational terms, e.g. 'mechanic', are closer to words representing men ('man', 'boy', 'he', etc.), while other terms, e.g. 'nurse', are closer to words representing women ('woman', 'girl', 'she', etc.), indicating a gender bias \cite{garg2018word}. By training separate word embedding models on time-segmented historical texts, it has been further demonstrated that the changes in word distances over time reflect real-world changes in women’s occupations.

Another study has found that words representing men are closer to words related to intelligence, while words representing women are closer to 'studying', reflecting a common stereotype in education: "boys are successful at school because they are smart and girls because they study a lot" \cite{boutyline2023school}. Training separate word embedding models on texts produced at different time points further showed that this stereotype emerged at specific point in time, consistent with sociological explanations of the phenomenon.

Simply computing word similarities in a given corpus is often not very informative. Therefore, researchers typically segment the corpus for comparative analysis. These segments might represent different time periods, as in the examples above, or the corpus could be split by other criteria, such as training distinct models on texts authored by Republicans versus Democrats \cite{rodriguez2022word}. This approach, however, has a disadvantage as it reduces the amount of data available for training individual models, which could impair their performance. It also requires the alignment of resulting models in a common space, which could complicate the interpretation of the results \cite{hamilton2016diachronic}. 

In our work, we build upon these ideas by enhancing word embeddings with socio-demographic information and focusing on studying the self. More specifically, we replace every occurrence of the word 'I' in a large corpus of social media posts with $\text{I}_{g,a}$ tokens, where $g$ represents the gender of the post author and $a$ their age. We then train a word embedding model on the altered corpus. Projecting the resulting enhanced vectors on semantic axes allows exploring identity as expressed in social media posts. By incorporating socio-demographic information into the I-tokens, it also becomes possible to compare different social groups without splitting the original corpus.

In the remainder of this paper, we provide a more detailed description of our method. We then characterise the obtained enhanced I-tokens and verify whether they meet the criteria for face validity. To further validate our approach, we check if it can reproduce established findings on gendered self-views. We investigate the sensitivity of results with respect to model specifications and corpus size. Finally, we discuss the applicability of our approach to different settings and compare it with existing methods.

In the remainder of this paper, we provide a more detailed description of our method. We then characterise the obtained enhanced I-tokens and verify whether they meet the criteria for face validity. To further validate our approach, we check if it can reproduce established findings on gendered self-views. Next, we investigate the robustness of the results with respect to model specifications and corpus size. Finally, we discuss how our approach can be applied in different contexts and compare it with existing methods.

\section{Methods}
\subsection{Data \& Model}
To train the model, we used data on 62,707,791 posts shared over a span of 5 years by 913,230 users on the social media platform VK\footnote{\url{https://vk.com}}. Unlike on many other social media platforms, age and gender are mandatory fields of a user profile on VK and are publicly available via its API. This allows us to construct $\text{I}_{g,a}$ tokens for all posts in the dataset. While we use VK data for the results described in this paper, our approach could equally be applied to other data sources and to attributes beyond gender and age (see Discussion).

\begin{figure*}[!htb]
\centering
\includegraphics[width=\textwidth]{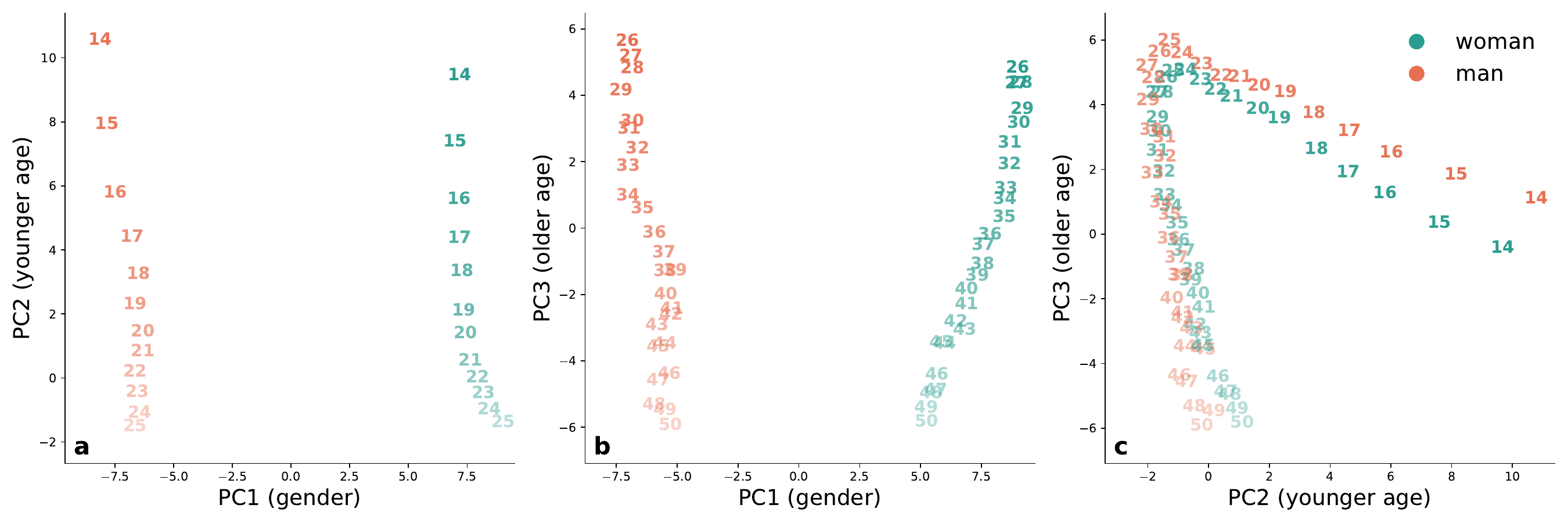}
\caption{\textbf{The structure of enhanced I-token embeddings.} The first principal component extracted from embeddings of enhanced I-tokens corresponds to gender (a, b). Curiously, age is represented by two components: the second component corresponds to a younger age (a), while the third corresponds to an older age (b).}
\label{fig:first}
\end{figure*}

We normalized all adjectives and nouns in the corpus using pymorphy2, the state-of-the-art morphological analyzer and generator for Russian and Ukrainian languages \cite{pymorphy}. This step is necessary because, in Russian, nouns and adjectives have distinct feminine and masculine forms. This makes words in feminine form artificially closer to $\text{I}_{\text{woman}, *}$ tokens in vector space and words in masculine form closer to $\text{I}_{\text{man}, *}$, preventing meaningful comparisons.

Next, we replaced all singular first-person pronouns used in posts with $\text{I}_{g,a}$ tokens, where $g$ and $a$ correspond to the self-reported gender of an author and their self-reported age at the time of writing. We then trained a continuous bag-of-words model \cite{mikolov2013efficient} with 100 dimensions over 10 epochs on this modified corpus. We report the main results for this specific model configuration; however, we also examine their sensitivity to model type (CBOW vs skip-gram), number of dimensions, number of epochs, and corpus size.

We examined the geometric structure of the obtained enhanced embeddings to ensure their face validity. Specifically, we expect $\text{I}_{\text{men},*}$ and $\text{I}_{\text{woman},*}$ to be clearly separated in vector space. We also expect that $\text{I}_{g,a}$ tokens will be sequentially ordered by age, i.e., that $\text{I}_{g,i}$ would be between $\text{I}_{g,i - 1}$ and $\text{I}_{g,i + 1}$.

\textbf{Gendered self-views}

To further validate our approach, we checked if it can reproduce existing findings on sex-trait stereotypes. Sex-trait stereotypes refer to the psychological characteristics or behavioral traits believed to be more prevalent in women than in men, or vice versa \cite{williams1990measuring}. A common way to assess sex-trait stereotypes is to present participants with a series of adjectives and ask them to determine whether each adjective is more commonly associated with women or men. From such studies emerged a list of adjectives that participants consistently associate more with either women or men, whether they are describing others or themselves. Examples include 'affectionate' and 'sensitive' for women, and 'courageous' and 'ambitious' for men (the full list of adjectives used in this study is available in Table 1.1 of \cite{williams1990measuring}).

We translated this list into Russian and constructed a semantic axis (\textit{gender stereotype axis}) by subtracting the average embedding for men-associated adjectives from the average embedding for women-associated adjectives. While the original list was obtained by asking Euro-American college students, recent studies demonstrate that women and men consistently rate themselves higher on corresponding traits across 62 countries \cite{kosakowska2023gendered}. Thus, if our approach is valid, we expect the projections of $\text{I}_{\text{woman},*}$ on the \textit{gender stereotype axis} to be positive, while projections of $\text{I}_{\text{man},*}$ to be negative.

\section{Results}

We found that the variation between $\text{I}_{g,a}$ tokens is largely explained by gender and age variables. In particular, the first principal component extracted from these vectors corresponds to gender, clearly separating $\text{I}_{woman, *}$ from $\text{I}_{man, *}$ tokens (Figure \ref{fig:first}a and \ref{fig:first}b). The point-biserial correlation coefficient between gender and the first component is $0.986$ ($P < 10^{-61}$).

\begin{figure*}[!htb]
\centering
\includegraphics[width=\textwidth]{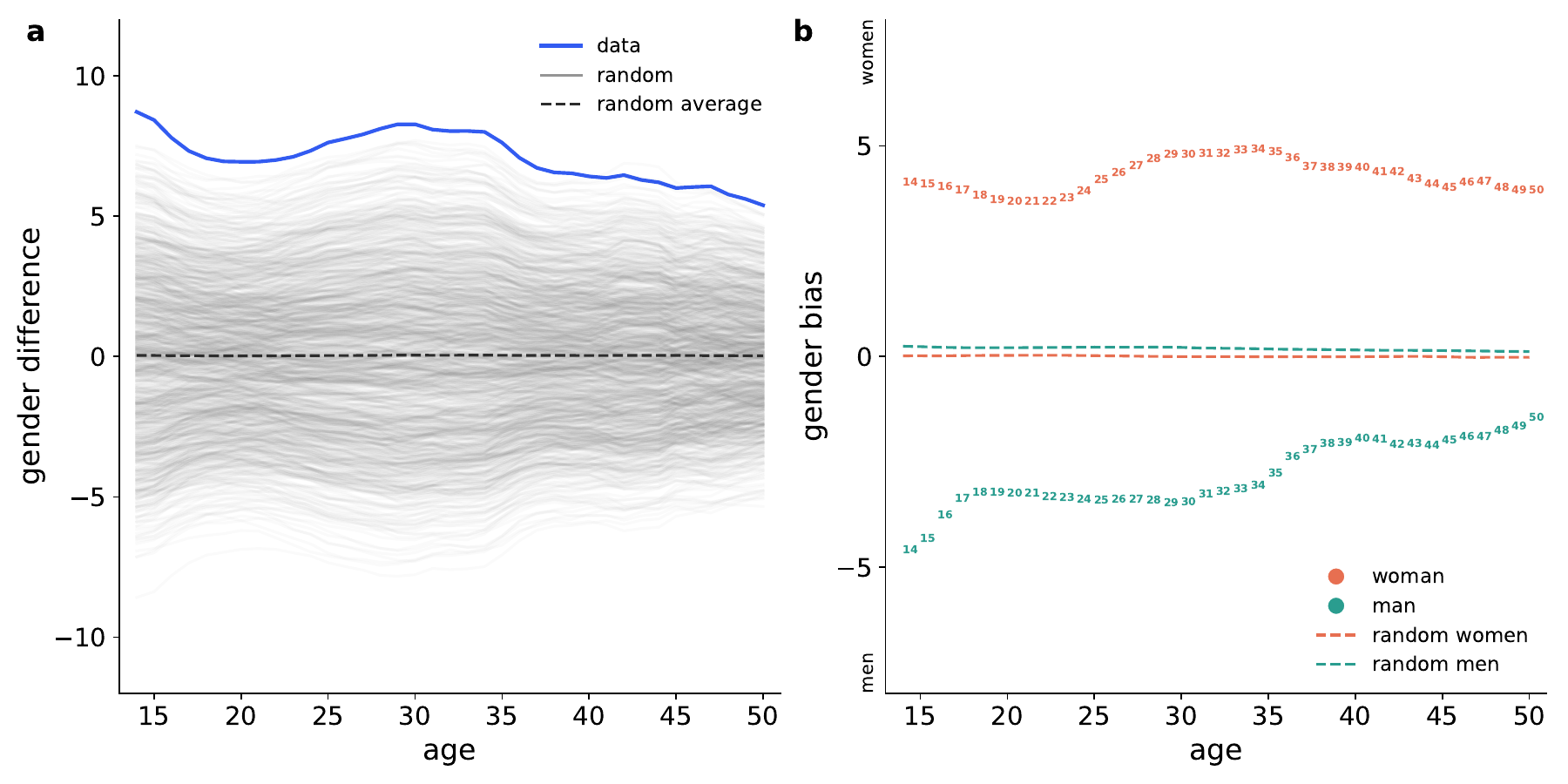}
\caption{\textbf{Projection of enhanced I-tokens on \textit{gender stereotype axis} reproduces established findings on gendered self-views.} $\text{I}_{\text{woman},*}$ tokens are closer to women's pole of the axis, while $\text{I}_{\text{man},*}$ tokens are closer to the men's pole, with the distance between them being larger than what could be explained by chance (a). The gap narrows with age as $\text{I}_{\text{man},*}$ tokens shift towards the center (b).}
\label{fig:second}
\end{figure*}

We expected that the second principal component would correspond to age. However, the results are more nuanced: the second component corresponds to younger age (Figure \ref{fig:first}a) with Spearman's $\rho = 0.965, P < 10^{-13}$, while the third component corresponds to older age (Figure \ref{fig:first}b) with Spearman's $\rho = 0.928, P < 10^{-24}$. The interaction between these components and age is shown in Figure \ref{fig:first}c. We hypothesise that this might be explained by graduation from university and the transition to working life, as the curve's turning point (25--26 years in Figure \ref{fig:first}c) roughly matches the age when students typically complete their degrees in Russia.

\begin{figure*}[!htb]
\centering
\includegraphics[width=\textwidth]{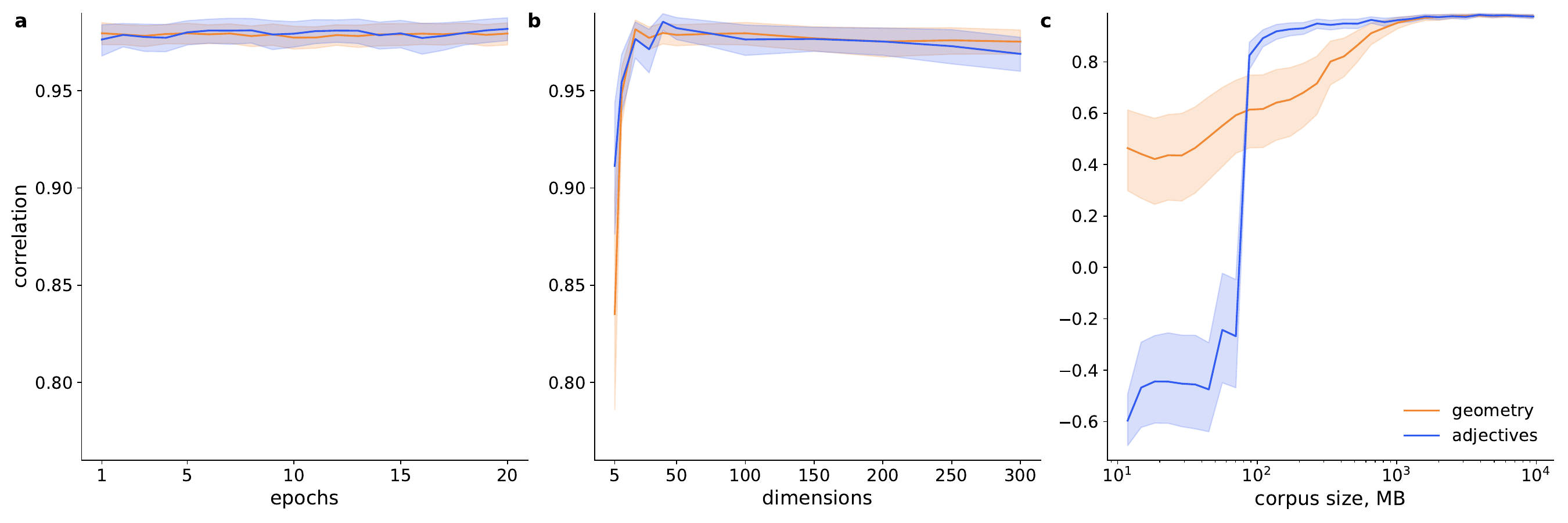}
\caption{\textbf{Robustness of the results with respect to model specification.} We evaluated how much point-biserial correlations between gender and the first principal component extracted from enhanced I-tokens (orange), as well as between gender and the projection of I-tokens on the \textit{gender stereotype axis} (blue), depend on model specification. We found that no further training is required beyond one epoch to reproduce the results (a). We also found that any reasonable number of dimensions can be used (b). Finally, we found that 100MB is a sufficient corpus size, but beyond that point, the performance drops for adjectives as they become too rare. The first principal component of enhanced I-tokens remains strongly associated with gender for all our experiments.
}
\label{fig:third}
\end{figure*}

\textbf{Gendered self-views}

If our approach is valid, we expect that projections of $\text{I}_{\text{woman},*}$ on the \textit{gender stereotype axis} will be positive, and projections of $\text{I}_{\text{man},*}$ will be negative. This is indeed what we observe (see Figure \ref{fig:second}b). We tested the significance of this result by randomly shuffling adjectives used to construct the \textit{gender stereotype axis} and projecting the enhanced I-tokens on the resulting random axes (Figure \ref{fig:second}a). This makes these results significant with $P < 10^{-3}$.

Additionally, we were able to detect changes in the strength of this relationship over the years--a result that is difficult to capture in surveys, as they are typically conducted on samples of university students \cite{williams1990measuring,kosakowska2023gendered}. This demonstrates the potential of our methodology not only to reproduce established findings but also to gain new insights that might be harder to obtain via traditional methods.

\textbf{Robustness of the results}
We evaluated how well the observed geometrical structure of enhanced I-tokens is preserved across different model specifications. We also checked for the robustness of the relationship between gender and gender-stereotypical adjectives. For this purpose, we computed point-biserial correlation coefficients between gender and the first principal component extracted from I-tokens, as well as between gender and the projection of I-tokens on the \textit{gender stereotype axis} (see Figure \ref{fig:third}).

Although we trained the model for 10 epochs for the reported results, we found that further training offers little additional benefit and the main results can be reproduced after just one epoch (Figure \ref{fig:third}a). We also found that any reasonable number of dimensions (50--300) could be used without compromising the model's performance (Figure \ref{fig:third}b). The observed relationships are even more salient when vector sizes are between 50 and 100, which could be preferable due to the smaller model size. There was no substantial difference between the CBOW and skip-gram architectures.

We found that 100MB of data is sufficient to reproduce the results after training for one epoch (Figure \ref{fig:third}c). For smaller datasets, performance drops for adjectives because they become too rare in the corpus. The first principal component of enhanced I-tokens is strongly correlated with gender in all our experiments. In practice, an even smaller corpus could be used. For example, the corpus of interest could be augmented by a neural one, such as a Wikipedia dump. This should result in better representations of rare words without affecting enhanced I-tokens, as they would only be present in the original corpus of interest.

\section{Discussion}
In this paper, we introduced a novel approach that leverages readily available data sources, such as social media, to study identity. Unlike traditional methods that rely on self-report surveys, our method allows for the study of identity in natural settings and on a larger scale. While we used data from VK, the same technique can be applied to other datasets as well. For example, self-reported gender and age have been extracted from posts on popular platforms such as Reddit and Twitter \cite{tigunova2020reddust,klein2022reportage},  making it possible to apply our method directly to these datasets. Attributes that can be used to construct enhanced I-tokens are not limited to gender and age. For instance, with datasets containing profession information on Reddit \cite{tigunova2020reddust} or educational outcomes on VK \cite{smirnov2019schools}, it becomes possible to study differences between various socio-economic groups. Moreover, this approach can be extended beyond social media data. Our experiments demonstrate that the corpus does not need to be exceptionally large for the method to be effective. Therefore, it could be applied to TV scripts to analyse the representation of different groups on television, building upon previous research in this area \cite{ramakrishna2015quantitative,ramakrishna2017linguistic}.

As a proof of concept, we applied our method to study gendered self-views. We found that the approach not only reproduces established results but also allows for new findings by covering a wider age range than is typically available in surveys. This method can similarly be applied to other phenomena using curated word lists.  Alternatively, an open dictionary approach can be used to identify and examine words that are especially close to certain enhanced I-tokens in a corpus of interest.

The introduced method relies on natural language processing techniques that are admittedly no longer considered state-of-the-art. Since the introduction of word2vec \cite{mikolov2013efficient}, more advanced models have emerged, particularly fastText \cite{bojanowski2017enriching}, which operates at the character n-gram level and potentially offers superior embeddings for morphologically rich languages such as Russian. Later, contextual word embedding models were developed, most notably BERT \cite{devlin2018bert}, which outperformed static models in a wide range of tasks. However, we believe that advances in machine learning outpace their adoption in social sciences, and there are still many opportunities for new insights to be obtained from using static continuous representations of words. While newer models have led to remarkable performance gains in machine learning applications, we believe that the higher interpretability and computational efficiency of simpler models might still make them preferable for analytical purposes and applications in social science.
 
The idea of using semantic projections traces its origins back to at least 2016 \cite{bolukbasi2016man}, when a gender axis was constructed to reveal biases in word embeddings. This methodology was later formally introduced in \cite{an2018semaxis}, re-introduced in \cite{mathew2020polar}, and re-introduced again in \cite{grand2022semantic}. It was further extended to contextual word embeddings \cite{lucy2022discovering,engler2022sensepolar}. Despite these developments within the computer science literature, their adoption in social sciences has been relatively slow. One possible explanation is that these methods enable the identification of biases at an aggregated level of entire corpora, which, while interesting, has limited applications. In our paper, we build upon previous ideas and show how they can be extended to study differences between social groups. We believe this opens up many new possibilities that would be particularly appealing to social scientists.

\bibliography{paper}

\end{document}